# Refinement of Uncertainty Relations in Quantum Mechanics

## Sergei P. Efimov

Bauman Moscow State Technical University,
2-ya Baumanskaya ul. 5, str.1, 105005  Moscow, R F

## Abstract
The uncertainty relation of three quantities in quantum mechanics is estimated in terms of commutators. The Pauli matrices are used to find a contribution of third-order commutators. The resulting inequality refines the Heisenberg indeterminacy for two non-conjugate operators when their commutator is not a c-number. The inequality is applied to operators of the kinetic energy and coordinate of one-dimensional system. In addition, uncertainty of the angular momentum components is considered.

## I. INTRODUCTIAN

The uncertainty relation or "the Heisenberg indeterminacy" relation in quantum mechanics is a consequence of commutative properties of the quantities [1-3]. Their operators has been properly used for the construction of the "most classical" states [5]. It appears that a minimum of the uncertainties of two operators do not always lead to a coherent state.

According to Jackiw[5], to construct a state with minimum of uncertainties, we have to consider inequality for two deviations of the Hermitian operators $\hat{A}$ and $\hat{B}$

$$\langle (\delta\hat{A})^2 \rangle \langle (\delta\hat{B})^2 \rangle \geq \frac{1}{4}(\hat{C})^2, \qquad (1)$$

where deviation of operator from its average is noted here as $\delta\hat{A} = \hat{A} - \langle \hat{A} \rangle$ and $\hat{C}$ is the commutator

$$\hat{C} = i[\hat{A}, \hat{B}].$$

As a rule, we have in Eq. (1) symbol >, i.e., quantity on the left-hand side more than one on the right-hand. When the operator $\hat{C}$ is c-number only, Eq. (1) can be the equality to give the minimum strictly. Found state yields then coherent state. To be precise, as mathematical result, Eq. (1) is the Robertson uncertainty. History of issue presented in [1].

In the general case, to minimize the product of the uncertainties, we have to solve equation

$$\left( \frac{(\delta\hat{A})^2}{\langle (\delta\hat{A})^2 \rangle} + \frac{(\delta\hat{B})^2}{\langle (\delta\hat{B})^2 \rangle} - 2 \right) |\Psi\rangle = 0 \qquad (2)$$

analytically. The minimum of the uncertainties product cannot then be expressed in simple terms of operators $\hat{A}$ and $\hat{B}$ ( like $\langle C \rangle^2$ ).

Further, we find an inequality that refines Eq. (1). It is constructed for several, pairwise non-commuting operators. So, we are not looking for minimum of uncertainties but refine the valuation in terms of third-order commutators.

## II. The PAULI MATRICES in INEQUALITY

When a commutator $\hat{C}$ is nearly a c-number, i.e., its fluctuations are small, the minimum of the left-hand side of Eq. (1) is $\langle C \rangle^2$ approximately. However, minimization of the commutator variance contradicts to the minimization of the uncertainties. Using commutators of third-order, we have inequalities:

$$\langle \hat{C}^2 \rangle \langle (\delta \hat{A})^2 \rangle \geq \frac{1}{4} \langle i[\hat{A},\hat{C}] \rangle^2, \tag{3}$$

$$\langle \hat{C}^2 \rangle \langle (\delta \hat{B})^2 \rangle \geq \frac{1}{4} \langle i[\hat{A},\hat{C}] \rangle^2. \tag{4}$$

A compromise between the requirements of Eqs. (1), (3), and (4) can be made by an inequality in which commutators of the third-order are used. Some inequality is asking for summing Eqs. (1), (3), and (4). We can assert, however, that the inequality does not become an equality at any case to give the minimum.

We obtain a relation by using a quadratic form whose coefficients are determined by the fluctuations of the operators $\hat{A}$, $\hat{B}$ and their commutators. For this purpose, we consider the Pauli $2 \times 2$ matrices $\hat{\boldsymbol{\sigma}}_i$ to put them in an operator:

$$\hat{F} = \gamma_1 \delta \hat{A} \otimes \hat{\boldsymbol{\sigma}}_x + \gamma_2 \delta \hat{B} \otimes \hat{\boldsymbol{\sigma}}_y + \gamma_3 \hat{C} \otimes \hat{\boldsymbol{\sigma}}_z, \tag{5}$$

where constants $\gamma_i$ are taken real.

The squared operator is equal to

$$\hat{F}\hat{F}^+ = \gamma_1^2 (\delta \hat{A} \delta \hat{A}^+) + \gamma_2^2 (\delta \hat{B} \delta \hat{B}^+) + \gamma_3^2 (\hat{C}\hat{C}^+) + \gamma_1 \gamma_2 \hat{C} \otimes \hat{\boldsymbol{\sigma}}_z + \gamma_2 \gamma_3 \hat{C}_2 \otimes \hat{\boldsymbol{\sigma}}_x - \gamma_1 \gamma_3 \hat{C}_3 \otimes \hat{\boldsymbol{\sigma}}_y, \tag{6}$$

where $\hat{C}_2$ and $\hat{C}_3$ are commutators of the third-order:

$$\hat{C}_2 = i[\delta \hat{B}, \hat{C}], \tag{7}$$

$$\hat{C}_3 = i[\delta \hat{A}, \hat{C}]. \tag{8}$$

Operator from Eq. (6) is positive-definite. We average the equation over the coordinate state $|\psi\rangle$, obtaining an operator in the space of "spin" variables:

$$\langle \Psi | \hat{F}\hat{F}^+ | \Psi \rangle = \gamma_1^2 \langle (\delta \hat{A})^2 \rangle + \gamma_2^2 \langle (\delta \hat{B})^2 \rangle + \gamma_3^2 \langle \hat{C}^2 \rangle + \gamma_1 \gamma_2 \langle \hat{C} \rangle \otimes \hat{\boldsymbol{\sigma}}_z + \gamma_2 \gamma_3 \langle \hat{C}_2 \rangle \otimes \hat{\boldsymbol{\sigma}}_x - \gamma_1 \gamma_3 \langle \hat{C}_3 \rangle \otimes \hat{\boldsymbol{\sigma}}_y \tag{9}$$

Since the parameters $\gamma_i$ are arbitrary, we arrive at the positive-definite matrix $6 \times 6$, that determinant is of :

$$\text{det} = \begin{vmatrix} \langle(\delta\hat{A})^2\rangle\hat{\mathbf{1}} & \frac{1}{2}\langle\hat{C}\rangle\hat{\sigma}_z & -\frac{1}{2}\langle\hat{C}_3\rangle\hat{\sigma}_y \\ \frac{1}{2}\langle\hat{C}\rangle\hat{\sigma}_z & \langle(\delta\hat{B})^2\rangle\hat{\mathbf{1}} & \frac{1}{2}\langle\hat{C}_2\rangle\hat{\sigma}_x \\ -\frac{1}{2}\langle\hat{C}_3\rangle\hat{\sigma}_y & \frac{1}{2}\langle\hat{C}_2\rangle\hat{\sigma}_x & \langle\hat{C}^2\rangle\hat{\mathbf{1}} \end{vmatrix}$$

. (10)

Principal minors of it are non-negative. The unit matrix $2\times 2$ is denoted as $\hat{\mathbf{1}}$. The Robertson uncertainty results from the upper minor of the second order that is non-negative.

To deduce general inequality, we substitute matrices

$$\hat{\sigma}_x = \begin{pmatrix} 0 & 1 \\ 1 & 0 \end{pmatrix}, \quad \hat{\sigma}_y = \begin{pmatrix} 0 & -i \\ i & 0 \end{pmatrix}, \quad \hat{\sigma}_z = \begin{pmatrix} 1 & 0 \\ 0 & -1 \end{pmatrix}$$

into Eq. (10). Lengthy calculation of determinant from Eq. (10) yields a product
$$M_1 M_2,$$
where first factor is

$$M_1 = \langle(\delta\hat{A})^2\rangle(\langle(\delta\hat{B})^2\rangle\langle\hat{C}^2\rangle - \frac{\langle\hat{C}_2\rangle^2}{4}) \\ + \langle(\delta\hat{B})^2\rangle(\langle(\delta\hat{A})^2\rangle\langle\hat{C}^2\rangle - \frac{\langle\hat{C}_3\rangle^2}{4})$$

(11)

and second one

$$M_2 = \begin{vmatrix} \langle(\delta\hat{A})^2\rangle & -\frac{1}{2}\langle\hat{C}\rangle & -\frac{i}{2}\langle\hat{C}_3\rangle \\ -\frac{1}{2}\langle\hat{C}\rangle & \langle(\delta\hat{B})^2\rangle & \frac{\langle\hat{C}_2\rangle}{2} \\ \frac{i}{2}\langle\hat{C}_3\rangle & \frac{\langle\hat{C}_2\rangle}{2} & \langle\hat{C}^2\rangle \end{vmatrix}$$

(12)

According to Eqs. (3) and (4) factor of Eq. (11) is nonnegative. We can reduce it to shorten Eq. (10) and leaving the factor $M_2$ only. By calculating now determinant of $3\times 3$ matrix, we arrive to relation

$$\langle(\delta \hat{A})^2\rangle\langle(\delta \hat{B})^2\rangle \geq \frac{1}{4}\left[\langle \hat{C}\rangle^2 + \frac{\langle \hat{C}_2\rangle^2 \langle(\delta \hat{A})^2\rangle}{\langle \hat{C}^2\rangle} + \frac{\langle \hat{C}_3\rangle^2 \langle(\delta \hat{B})^2\rangle}{\langle \hat{C}^2\rangle}\right]. \tag{13}$$

Eq. (13) is inequality that refines Heisenberg's indeterminacy when the commutator $\langle \hat{C}\rangle$ is not a c-number, i.e. $\langle \hat{C}_2\rangle$ or $\langle \hat{C}_3\rangle$ are non zero. We have to be careful in differing the quantities

$$\langle \hat{C}\rangle^2 \quad \text{and} \quad \langle \hat{C}^2\rangle.$$

Equality is valid only when an eigen-state of the operator $\hat{F}$ from Eq. (5) is realized to include the spin variables:

$$\hat{F}|\alpha\Psi\rangle = 0, \tag{14}$$

where $|\alpha\rangle$ is a spin state.

## III. UNCERTAINTY RELATION for KINETIC ENERGY and COORDINATE

We apply inequality of Eq. (13) to operators of kinetic energy $\hat{E}_{kin} = \dfrac{\hat{p}^2}{2}$ and coordinate $\hat{x}$. Operators are then following:

$$\langle(\delta \hat{A})^2\rangle = \langle(\delta \hat{E}_{kin})^2\rangle,$$
$$\langle(\delta \hat{B})^2\rangle = \langle(\delta \hat{x})^2\rangle,$$
$$\langle \hat{C}\rangle = \hbar\langle \hat{p}\rangle,$$
$$\langle \hat{C}^2\rangle = \hbar^2\langle \hat{p}^2\rangle = 2\hbar^2\langle E_{kin}\rangle$$
$$\hat{C}_2 = -\hbar^2,$$
$$\hat{C}_3 = 0.$$

Let the average impulse be not zero, i.e. $\langle \hat{p}\rangle \neq 0$. Substituting the quantities into inequality of Eq. (13), we obtain

$$\langle(\delta \hat{E}_{kin})^2\rangle\langle(\delta \hat{x})^2\rangle \geq \frac{\hbar^2}{4}\langle \hat{p}\rangle^2 + \frac{\hbar^2}{8}\frac{\langle(\delta \hat{E}_{kin})^2\rangle}{\langle E_{kin}\rangle}. \tag{15}$$

In particular, equality is achieved in Eq. (13) for the ground state of the oscillator whereas the right-hand part of the Robertson inequality vanishes. It is the essential refinement actually.

When $\langle \hat{p}\rangle$ is nonzero, Eq. (12) can be rewritten in the form

$$\langle(\delta \hat{E}_{kin})^2\rangle\langle(\delta t)^2\rangle \geq \frac{\hbar^2}{4} + \frac{\hbar^2}{4}\frac{\langle(\delta \hat{E}_{kin})^2\rangle}{\langle(\delta \hat{p})^2\rangle\langle \hat{p}\rangle^2}, \tag{16}$$

where

$$\langle(\delta t)^2\rangle = \frac{\langle(\delta \hat{x})^2\rangle}{\langle \hat{p}\rangle^2}$$

is the effective time to move particle near the average trajectory.

To estimate the product of the uncertainties, we use an inequality where neither $\langle(\delta\hat{A})^2\rangle$ nor $\langle(\delta\hat{B})^2\rangle$ do not arise on the right-hand of it. When the two last terms in Eq. (10) are of the same order, there is a reason to refine it. Regarding that, we use inequality

$$(a+b) \geq 2\sqrt{ab},$$

for positive values *a* and *b*. By continuing inequality of Eq. (13) by such way, Eq. (13) passes onto

$$\langle(\delta\hat{A})^2\rangle\langle(\delta\hat{B})^2\rangle\langle\hat{C}^2\rangle \geq \frac{1}{4}\langle\hat{C}\rangle^2\langle\hat{C}^2\rangle + \frac{1}{2}|\langle\hat{C}_2\rangle\langle\hat{C}_3\rangle|\sqrt{\langle(\delta\hat{A})^2\rangle\langle(\delta\hat{B})^2\rangle}. \tag{17}$$

Calculating a root of Eq. (17), we arrive to

$$\sqrt{\langle(\delta\hat{A})^2\rangle\langle(\delta\hat{B})^2\rangle} \geq \frac{1}{4\langle\hat{C}^2\rangle}\left[|\langle\hat{C}_2\rangle\langle\hat{C}_3\rangle| + \sqrt{\langle\hat{C}_2\rangle^2\langle\hat{C}_3\rangle^2 + 4\langle\hat{C}^2\rangle^2\langle\hat{C}\rangle^2}\right]. \tag{18}$$

### IV. UCERTAINTY RELATION for ANGULAR MOMENTUM

An inequality of the type of Eq. (10) can be constructed for several pairwise non-commuting operators. Let us consider as an example, the angular momentum (or spin) operators. We compile the operator

$$\hat{F} = \gamma_1(\delta\hat{L}_x)\otimes\boldsymbol{\sigma}_x + \gamma_2(\delta\hat{L}_y)\otimes\boldsymbol{\sigma}_y + \gamma_3(\delta\hat{L}_z)\otimes\boldsymbol{\sigma}_z. \tag{15}$$

We recall that the operators $\boldsymbol{\sigma}_i$ are auxiliary. Their commutative properties are of importance here only. As in Eq. (8), we may obtain a matrix that is positive-definite. Operators now are following:

$$\hat{C} = -\hat{L}_z, \quad \hat{C}_2 = \hat{L}_x, \quad \hat{C}_3 = -\hat{L}_y$$

$$\langle(\delta L_x)^2\rangle\langle(\delta L_y)^2\rangle\langle L_z^2\rangle \geq \frac{1}{4}\langle L_z^2\rangle\langle L_z\rangle^2 + \frac{\langle(\delta\hat{L}_x)^2\rangle\langle L_x\rangle^2}{4} + \frac{\langle(\delta\hat{L}_y)^2\rangle\langle L_y\rangle^2}{4} \tag{16}$$

### V. CONCLUSIONS

The method makes it possible to obtain other constraints. For this purpose, we must construct an operator analogous to Eq. (5) using commutators of higher orders, and replace the Pauli matrices by matrices forming a Clifford algebra [ 6 ]. The idea of [8] was developed in prospective studies [ 9,10 ].